\documentclass[aps,pra,preprint,amsmath,amssymb,superscriptaddress]{revtex4-1}
\usepackage{bm}
\usepackage{amssymb}
\usepackage{colordvi}
\usepackage{graphicx}
\usepackage{color}
\usepackage{hyperref}

\newcommand{\be}{\begin{equation}}
\newcommand{\ee}{\end{equation}}
\newcommand{\bea}{\begin{eqnarray}}
\newcommand{\eea}{\end{eqnarray}}

\newcommand{\vep}{\varepsilon}

\newcommand{\ome}{\omega}

\def\ket#1{\vert #1 \rangle}

\begin{document}

\title{Type-II Dirac Photons}

\author{Hai-Xiao Wang}\email{These authors contributed equally.}
\affiliation{College of Physics, Optoelectronics and Energy, \&
  Collaborative Innovation Center of Suzhou Nano Science and
  Technology, Soochow University, 1 Shizi Street, Suzhou 215006,
  China}
\author{Yige Chen}\email{These authors contributed equally.}
\affiliation{Department of Physics, University of Toronto, Toronto,
  M5S 1A7, Canada}
\author{Zhi Hong Hang}
\affiliation{College of Physics, Optoelectronics and Energy, \&
  Collaborative Innovation Center of Suzhou Nano Science and
  Technology, Soochow University, 1 Shizi Street, Suzhou 215006,
  China}
\author{Hae-Young Kee}\email{hykee@physics.utoronto.ca}
\affiliation{Department of Physics, University of Toronto, Toronto,
   M5S 1A7, Canada}
\affiliation{Canadian Institute for Advanced Research, Toronto, Ontario, M5G 1Z8, Canada}
\author{Jian-Hua Jiang}\email{jianhuajiang@suda.edu.cn}
\affiliation{College of Physics, Optoelectronics and Energy, \&
  Collaborative Innovation Center of Suzhou Nano Science and
  Technology, Soochow University, 1 Shizi Street, Suzhou 215006,
  China}

\date{\today}

\maketitle

\noindent {\bf {\Large Abstract}}\\
{\bf The Dirac equation for relativistic electron waves is the parent model for Weyl 
and Majorana fermions as well as topological insulators. Simulation of Dirac
physics in three-dimensional photonic crystals, though fundamentally important
for topological phenomena at optical frequencies, encounters
the challenge of synthesis of both Kramers double degeneracy and
parity inversion. Here we show how type-II Dirac points---exotic Dirac
relativistic waves yet to be discovered---are robustly realized
through the nonsymmorphic screw symmetry.
The emergent type-II Dirac points carry nontrivial topology
and are the mother states of type-II Weyl points. The proposed all-dielectric
architecture enables robust cavity states at
photonic-crystal---air interfaces and anomalous 
refraction, with very low energy dissipation.
}\\

\noindent {\bf \Large Introduction}\\
Dirac's famous equation for relativistic electron waves \cite{dirac} is the
foundation for both the quantum field theory and the later topological
insulators and semimetals \cite{ti1,ti2,tsm1}. There has been a
trend in the simulation of relativistic waves and topological states 
in classical dynamics such as electromagnetic \cite{haldane,rev1},
acoustic \cite{bzhang,acoustic,nju1} and mechanical
waves \cite{huber,vinzo}, mostly in 2D systems. Many novel
phenomena in electromagnetism are discovered along this paradigm, such
as photonic {\em Zitterbewugung} \cite{ZB}, zero-index dielectric 
metamaterials \cite{zim}, deformation induced pseudomagnetic field for
photons \cite{sMag}, as well as photonic topological
insulators with \cite{z2meta,ctti,shvets,huxiao,oe1} and  
without \cite{haldane,mit,wu,hafezi2,floquet,hafezi3} 
time-reversal (${\cal T}$) symmetry. Recently, such simulation
develops from 2D to
3D \cite{ling1,ling-exp,szhang,ct-exp,3ddp,xiao,3dti,3dwti},
exposing to larger wavevector and configuration space that may lead
to richer physical phenomena, particularly using ${\cal
 T}$-invariant materials which are more feasible for
high-frequency (e.g., infrared or visible) applications.

{Due to its bosonic nature, i.e., ${\cal T}^2=1$, the four-fold
degenerate photonic Dirac points (DPs) can be created only when
Kramers double degeneracy (``spin'') and parity-inversion (``orbit'')
are {\em synthesized}. These two elements are also at the heart of
$Z_2$ topology in ${\cal P}{\cal T}$-symmetric (${\cal P}$ is
inversion) systems, as revealed in the seminal work of Fu and Kane
\cite{fukane}. Although there have been a few fine 
designs \cite{3ddp,3dti,3dwti} showing the connection between type-I
DPs and the $Z_2$ topology, type-II DPs [{in analog of type-II Weyl
Points (WPs) \cite{WPII,WP2}}, see Fig.~1] have never been
explored in photonics or in other {classical/}bosonic waves. In this
work, we demonstrate the creation and destruction
of type-II DPs in PhCs. Besides, we unveil screw symmetry, a
fundamental type of nonsymmorphic symmetry, as an effective tool for
the creation of DPs.}

The distinction between symmorphic (e.g., point-group) and
nonsymmorphic spatial symmetries in crystals lies in whether the
spatial origin can be preserved. Nonsymmorphic symmetries cannot
preserve the spatial origin but translate it by a fraction of the
crystal period. {The screw symmetry, a rotation accompanied with a
fraction of lattice translation, is an elementary nonsymmorphic
symmetry.} So far, the role of screw symmetry on the realization of
topological states in classical/bosonic waves has not yet been
explored. {It is known that screw symmetries lead to double
degeneracy for all Bloch states on certain planes in the Brillouin
zone (BZ) \cite{mermin,ashvin}. Thus the screw symmetries can create a large
wavevector space for the simulation of DPs and $Z_2$ topology in
classical dynamics.} The screw symmetries become particularly powerful
when there are two orthogonal screw axes, since the product of the two
screw rotations is essentially the parity required by the DPs. In this
way both the ``Kramers'' double degeneracy (``spin'') and
parity-inversion (``orbit'') can be simultaneously synthesized through
screw symmetry.

{Based on these symmetry considerations we propose an all-dielectric
tetragonal PhC with screw symmetries for the creation of
both type-II and type-I DPs. Our symmetry-guided approach is
robust: DPs emerge for a variety of geometry and
materials. We demonstrate the nontrivial topology of the 
DPs by studying the edge states.} These non-chiral edge states,
differing from the chiral edge states of 
Weyl points (WPs), are below the light-line and form resilient
cavity states on PhC-air interfaces. Moreover, we show that both
type-II and type-I WPs can be derived from these DPs when symmetry is 
reduced. Anomalous refraction with one or two pairs of opposite
refraction angles is predicted for type-II DPs/WPs. To the best of our
knowledge, this is the discovery of type-II DPs in photonics and
a proposal of type-II WPs in {\em all-dielectric} PhCs. Our
findings may enable unprecedented control of light at optical
frequencies {using dissipationless materials}.\\

\noindent {\bf \Large Results}\\
\noindent {\bf All-dielectric photonic-crystal architecture}\\
{We study an all-dielectric PhC with tetragonal symmetry of space group
P4$_2$/mcm (see Fig.~2) to illustrate the symmetry-guided approach.
In each unit cell, there are two dielectric blocks (painted as yellow and
green in Fig.~2) of the same shape and permittivity $\vep_b$, embedded
in a polymer matrix of permittivity $\vep_m=1.9$. We shall first set
$\vep_b=16$ and the geometry parameters $l=0.5$, $w=0.2$, and $h=0.5$
(lattice constant $a\equiv 1$).
We show later that DPs emerge for other material/geometric parameters
as well.} These PhCs can in principle be fabricated
using layer-by-layer methods with the current technology \cite{ald,dlw}
for infrared frequencies. We use the MIT PHOTONIC BANDS \cite{mpb} to
calculate the bulk and surface photonic bands. {The tetragonal
symmetries crucial to our study are the 
two-fold screw symmetries $S_x := (x,y,z)\rightarrow (\frac{1}{2}+x, 
\frac{1}{2}-y, \frac{1}{2}-z)$ and $S_y := (x,y,z)\rightarrow
(\frac{1}{2}-x, \frac{1}{2}+y, \frac{1}{2}-z)$ (illustrated in
Fig.~2b), and the 180$^\circ$ rotation around the $z$ axis,
$C_2:=(x,y)\rightarrow (-x,-y)$.} The remaining
symmetries are listed and analyzed in the Supplementary Materials.\\


\noindent {\bf Photonic Kramers double degeneracy}\\
{Anti-unitary operators: $\Theta_i\equiv S_i* {\cal T}$ ($i=x,y$)
are created to elucidate the power of the screw symmetry. The effect
of the time-reversal operation ${\cal T}$ on a photonic Bloch
wavefunction $\Psi_{n{\vec k}}({\vec r})=({\vec e}_{n{\vec k}}, {\vec h}_{n{\vec
    k}})^T$ is mostly complex conjugation, ${\cal T}({\vec
  e}_{n{\vec k}}, {\vec h}_{n{\vec k}})^T=({\vec e}^\ast_{n{\vec k}},
-{\vec h}^\ast_{n{\vec k}})^T$.} Since $\Theta_x^2=S_x^2=T_{100}$ where $T_{100}$ is a spatial translation
by the vector $(1,0,0)$, acting $\Theta_x$ twice 
on a photonic Bloch state  gives
${\Theta}_x^2 \Psi_{n{\vec k}}({\vec r}) = e^{i k_x} \Psi_{n{\vec
    k}}({\vec r})$ (see details in Methods). $\Theta_x$ transforms
$(k_x, k_y, k_z)$ into $(-k_x, k_y, k_z)$ and is hence invariant
on the $k_x=\pi$ plane, {where we find}
\vskip -0.5cm
\begin{align}
{\Theta}_x^2=\left. e^{i k_x}\right|_{k_x=\pi}= -1 .
\end{align}
{The above equation, as an analog of the Kramers theorem for fermions,
guarantees that {\em all} photonic} states on the $k_x=\pi$ plane are
doubly degenerate (see Fig.~2c). Similarly,
all Bloch states are doubly degenerate on the $k_y=\pi$ plane due to $\Theta_y^2=-1$.\\

\noindent {\bf Dirac Points}\\
{For the creation of DPs, the next important step is to realize
parity-inversion. Here the parity is defined through $C_2$, which is
invariant on the MA line, $k_x=k_y=\pi$.} The product of the two
orthogonal screw rotations yields, $S_yS_x=T_{010}C_2$ and
$S_xS_y=T_{100}C_2$. {On the MA line one hence has}
\vskip -0.5cm
\vskip -0.5cm
\begin{align}
\Theta_y \Theta_x = \Theta_x \Theta_y = - C_2 .
\end{align}
Elegantly, the above algebra reveal that the two degenerate states in
any doublet have the {\em same} eigenvalue of the $C_2$ rotation. Such
eigenvalues $c_2=\pm 1$ precisely  represent the parities of the
photonic states in the $x$-$y$ plane. 

{It has been shown in Ref. \cite{3ddp} that a DP with synthetic
Kramers double degeneracy and parity-inversion has nontrivial topological
properties. In fact, such DPs are monopoles of the $SU(2)$ Berry-phase gauge
fields \cite{3ddp}. The topological charge of a DP is defined by the integral of
the {$SU(2)$} gauge fields over a
{tiny} sphere containing the DP. It was 
proved in Ref.~\cite{yang} that in systems with ${\cal P}{\cal T}$
symmetry, the calculation of the topological charge of a DP can be
simplified as}
\vskip -0.5cm
\begin{align}
N_{DP} = \frac{1}{2}\left[c_2^{-}(k_0^+) -
c_2^{-}(k_0^-)\right],
\end{align} 
where $c_2^{-}$ is the parity of the lower branch of the Dirac
cone, and $k_0^+=k_0+0^+$ ($k_0^-=k_0-0^+$) is the 
wavevector slightly larger (smaller) than that of the DP
on the $z$ direction, $k_0$. {Since the total topological charge of
photonic bands in the BZ is strictly zero, DPs emerge in pairs with
opposite $N_{DP}$ at opposite wavevectors.} Fig.~2d shows that there
are four DPs in the first six bands, due to the crossing between the
$p$- and $d$-doublets.

Our symmetry-guided paradigm provides a robust and effective approach
toward topological DPs: Fig.~2e shows that the emergence of DPs is
quite robust to the shape and permittivity of the dielectric blocks
(more examples are given in the Supplementary Materials),
{since any crossing between bands of different
  parities on the MA line leads to DPs.}

The spin-orbit physics of the Dirac points can be understood via 
a symmetry-based ${\vec k} \cdot{\vec P}$ theory (see Supplementary
Materials for details). The Hamiltonian can be constructed using the
basis of the two doublets, $p_1$, $p_2$, $d_1$ and $d_2$
[Fig.~2d]. {The combination of these states,}
$\ket{p_{\pm}}=\frac{1}{\sqrt{2}}(\ket{p_1}\pm i \ket{p_2})$ and  
$\ket{d_{\pm}}=\frac{1}{\sqrt{2}}(\ket{d_1}\pm i \ket{d_2})$,
carry {finite} total angular momenta (TAM) that are opposite for the $+$ and
$-$ states (see Supplementary Materials). {Emulating fermionic spin and
orbit with the TAM and parity, respectively, we find the following photonic
Hamiltonian for a DP,} 
\vskip -0.5cm
\begin{align}
\hat{{\cal H}} = \ome_0 + v \left( \begin{array}{ccccc}
    (\eta+1) q_z\hat{1} & \hat{{\cal A}} \\
    \hat{{\cal A}}^\dagger &  (\eta-1) q_z \hat{1} \\
  \end{array}\right) + {\cal O}(q^2) , \quad \hat{{\cal A}} \equiv
{ g_0
\hat{1} + {\vec g}\cdot\hat{{\vec \sigma}} }. \label{kp}
\end{align}
where $\ome_0$ is the frequency of the DP, $v$ is the
characteristic group velocity. $\hat{1}$ is the $2\times 2$ identity
matrix, $\hat{{\vec \sigma}}$ is the Pauli matrix vector. {The
dimensionless ${\vec k} \cdot{\vec P}$ parameter $\eta$ here plays an
role to distinguish the type-I  ($|\eta|<1$) and type-II ($|\eta|>1$)
DPs. $g_0=i\alpha q_y$, $g_x=-i\alpha q_x$, $g_y=i\beta q_x$, $g_z=
-\beta q_y$ with ${\vec q}\equiv {\vec k}-(\pi,\pi,k_0)$, where $\alpha$ and $\beta$ are the
(real) ${\vec k}\cdot{\vec P}$ coefficients}, and ${\cal O}(q^2)$
denotes the higher-order quadratic warping terms. {Here the
spin-orbit coupling is emulated by the $k$-linear interaction between
the $p$ and $d$ bands due to quasi-conservation of the TAM \cite{3ddp}. The 3D
Dirac wave can be regarded as a series of $q_z$-dependent 2D Dirac
waves of which the Dirac mass, $m_D\equiv vq_z$, can be positive,
negative, or zero \cite{furusaki,3ddp}.}\\

\noindent {\bf Derived type-II and type-I Weyl Points}\\
A DP can be regarded as composed of a pair of WPs of opposite Chern
numbers. Thus when the space symmetry is reduced WPs can emerge from
DPs \cite{3ddp}. To realize the WPs, we deform the unit-cell
structure in such a way (as displayed in Fig.~3a) that the two screw
symmetries $S_x$ and $S_y$, the three mirror symmetries $M_1 :=
(x,y)\rightarrow(y,x)$, $M_2 := (x,y)\rightarrow (-y,-x)$, and $M_z:=
z\rightarrow -z$, as well as the inversion symmetry ${\cal P}$ are
broken. However, the $C_2$ symmetry is  preserved. The removal of
the two screw symmetries lifts the  double degeneracy on
the MA line. However, accidental degeneracy between bands of opposite
parity is protected by the $C_2$ symmetry. The chiral structure of the
PhC results in $p_{\pm}$- and $d_{\pm}$-like states in the photonic
bands. The crossings between the $p$ and $d$ bands results in WPs of
Chern number $\pm 1$ (see Supplementary Materials for a ${\vec
  k}\cdot{\vec P}$ analysis). We identify six WPs in Fig.~3b (there
are more WPs at higher frequency, explaining the nonzero total
Chern number). Fig.~3b also shows that there are four type-II WPs and two
type-I WPs. The 3D dispersions of both type-I and type-II WPs on the
lowest $d$-band are shown in Fig.~3c. {Our PhC
  architecture thus allows realization of type-II WPs using
  dissipationless all-dielectric materials.}\\

\noindent {\bf Robust surface states}\\
According to the bulk-edge correspondence principle \cite{ti1,ti2,tsm1},
the (100) surface states of the tetragonal PhC can reveal the $Z_2$
topology of the DPs. We then calculate the surface and projected bulk
photonic spectrum using a supercell stacking along the $x$ direction
[see Methods]. Fig.~4a shows a gapless surface band traversing the
projected photonic band gap. This surface band is between the
upper and lower branches of the type-I DP, but above both branches of
the type-II DP. Thus the gapless surface band is induced by the type-I
topological DPs. Nonetheless, both type-I and type-II DPs have the
same $Z_2$ topology (see Fig.~2d). {The topological surface states
carry finite TAM as indicated in Fig.~4b by the winding profile of the
Poynting vectors. The sign of the photonic TAM is changed when the
wavevector is reversed (see Fig.~4b). This property} is similar to the 
``spin-wavevector locking'' on the edges of topological
insulators \cite{ti1,ti2}. We find that the two
symmetries, $S_y$ and ${\cal T}$, guarantee that the
spectrum in the surface BZ is symmetric under the transformation
$(k_y,k_z) \to (\pm k_y, \pm k_z)$ (see Methods). It was recently
discovered that the surface states of the topological DPs form a
double-helicoid surface states with such spectral symmetry. The
non-chiral surface bands of our PhC, are distinctive from the chiral
surface states due to WPs \cite{ling1,ct-exp,szhang}.
Moreover, the topological surface states here are
below the light-line and hence form cavity states on the PhC-air
interfaces {with no need for} additional cladding.

The robustness of the topological surface states can be revealed via
their frequency stability against surface modifications. Fig.~4c shows
that the frequency of the topological surface state is quite robust
and insensitive to variations of the thickness of a dielectric slab
placed on top of the PhC surface. The change of frequency is within
2.5\%, although the field profile has been substantially modified
(see Fig.~4d). {In contrast, the frequency of a conventional PhC cavity
state with woodpile-PhC cladding is much more sensitive to the
thickness of the slab \cite{sjyang} (see inset of Fig.~4c and details
in Methods), despite the fact that the woodpile PhC
has a large complete photonic band gap of $\delta\ome/\ome=21\%$ while
our PhC has no complete photonic band gap}. The
topological surface states thus form resilient, subwavelength quasi-2D
photonic systems. The nontrivial topology/Berry-phases and the gapless
spectrum distinct them from normal PhC surface states \cite{noda2,book}.\\

\noindent {\bf Spectral and optical properties}\\
Both type-I and type-II DPs appear in Fig.~2. A more careful study is
presented in Fig.~5. From Eq.~(\ref{kp}), the spectrum of the DPs in
the $k_x$-$k_z$ plane (Fig.~5a) is 
\vskip -0.5cm
\begin{align}
\ome = \ome_0 + v \eta q_z + v\tau \sqrt{q_z^2 + \gamma^2
    q_{x}^2 }  ,\label{dpII}
\end{align}
where $\tau=\pm$ stands for the upper and lower branches of the DP,
respectively, the dimensionless parameters
$\gamma=\sqrt{\alpha^2+\beta^2}$ and $\eta$
measure the deformation of the Dirac cone. {Particularly,
$|\eta|>1$ for type-II DPs, whereas $|\eta|<1$ for type-I 
DPs.} The isofrequency contour near a type-II DP is a hyperbolic curve
(Fig.~5b). In contrast, the isofrequency contours near a type-I DP is
of elliptical shapes. For a type-II DP, when $\ome=\ome_0$, the two
branches touch each other and the isofrequency contour becomes a pair
of crossing lines (Fig.~5c), between which the angle is $\theta_{DP} = 2
\arctan\left(\sqrt{\frac{\eta^2-1}{\gamma^2}}\right)$. {This 
  quantity sets the bounds on the refraction angles near a type-II
  DP as $\pm \frac{1}{2}(\pi-\theta_{DP})$.}

{The dispersion of the type-II DP in the $k_x$-$k_y$
  plane is distinctive from the existing type-I DPs
  \cite{3ddp,3dti,3dwti} (see Fig.~5d). This spectrum can be
  understood via the ${\vec k}\cdot{\vec P}$ Hamiltonian (\ref{kp})
  which yields} $\ome_{\tau,i}({\vec q}) = 
\ome_0 + v \eta q_z + v\tau \sqrt{q_z^2 + \gamma^2 |q_i|^2 } + {\cal
  O}(q^2)$ for $i = 1, 2$, with $\tau=\pm$ and {$q_1=q_x+q_y$ and
$q_2=q_y-q_x$.}
This spectrum is nondegenerate for finite $q_x$ and $q_y$.
The two-fold degeneracy is restored only when {$q_x=0$ or $q_y=0$, in
accordance with the screw symmetries.} The ``V-shaped'' dispersion in
Fig.~5d gives elliptical-shaped isofrequency contours or non-closing
contours in the $k_x$-$k_y$ plane (see Figs.~5e and 4f), depending on
the quadratic warping terms.

{From the unique spectral properties of the type-II
  DPs, using frequency and wavevector matching, we derive the
  anomalous refraction of light: there are two concurrent refraction
  beams of opposite angles (see schematic in Fig.~5g). An analytic
  proof is detailed in the Methods section, which is confirmed by the
  model calculation in Figs.~5h and 5i for various frequencies,
  incident angles, and parameters. Interestingly, we find that there
  is no refraction for $\eta>1$, whereas for $\eta<-1$ there
  are two refraction beams of opposite refraction angles. Since the
  two DPs at opposite wavevectors have opposite $\eta$, the 
  above property can be exploited for {\em selective excitation} of
  type-II DPs.} Away from the $k_x=\pi$ and $k_y=\pi$ planes, the
photonic spectrum is nondegenerate, leading to two pairs of beams
with opposite refraction angles, as shown in Fig.~5i by varying the
angle of incidence $\phi_i={\rm Arg}(q_x+iq_y)$. Zero refraction
angle is realized when $\phi_i$ is close to $\frac{\pi}{4},
\frac{3\pi}{4}, \frac{5\pi}{4}$, or $\frac{7\pi}{4}$, due to vanishing
group velocity in the $k_x$-$k_y$ plane.

The above unconventional optical
properties also holds for type-II WPs. Since WPs are two-fold
degenerate, there can only be one pair of refraction beams. {Concurrent positive
and negative refraction was found and confirmed by time-domain
simulation in a 2D photonic system before \cite{luo}. Here we find,
from frequency-wavevector conservation, that concurrent negative and
positive refraction can also be realized in 3D all-dielectric PhCs
through type-II DPs/WPs. A time-domain simulation is demanded to
further investigate the anomalous refraction, which, however, is
beyond the scope of this work.}\\

\noindent {\bf \Large Discussion}\\
The band topology induced by crystalline symmetries are
in the context of topological crystalline states \cite{tci,ave}. Weak
disorders that preserve the crystalline symmetry on average should
preserve the DPs and their topological surface states
\cite{ave,3ddp}. {The topological surface states
  here can exist on the PhC-air interface without further cladding,
  even though such interface does not preserve the screw
  symmetries. The robustness of the surface photonic bands show
  superiority over conventional PhC cavity states. This suggests that 
  topology can be a possible tool to suppress inhomogeneous
  broadening which is a main obstacle for scalable optical and quantum
  devices.} Our all-dielectric topological PhC architecture may
inspire future discovery of other 3D topological photonic states in
all-dielectric photonics, and stimulate future synergy between
subwavelength photonic topological materials and optoelectronics
on PhC surfaces.\\

{
\noindent {\sl Note added}: When this paper was under review for the
final round, a proposal of type-II DPs in electronic materials with
robust Fermi arcs, has appeared \cite{hasan}.}\\

\noindent {\bf \Large Methods}\\
\noindent {\bf Symmetry transformation of photonic states}\\
A photonic state $\Psi_{n{\vec k}}({\vec r})$ transforms under the
$\Theta_x=S_x* {\cal T}$ operation as follows,
\vskip -0.5cm
\begin{align}
\Theta_x \Psi_{n{\vec k}}({\vec r}) =
\hat{M}_y\hat{M}_z\hat{t}_h\Psi^{\ast}_{n{\vec k}}(S_x{\vec r}) ,
\end{align}
where $\hat{M}_y$ and $\hat{M}_z$ are the mirror transformation for
the electromagnetic fields along the $y$ and $z$ directions, respectively. For instance,
\vskip -0.5cm
\begin{align}
& \hat{M}_y E_i = \bar{\delta}_{iy} E_i, \quad \hat{M}_z H_i = -
\bar{\delta}_{iz} H_i, \quad i = x,y,z ,\\
& \bar{\delta}_{ij}=\left \{\begin{array}{cccc} & 1 , \quad {\rm if}\
    \ 
    i\ne j\\ & -1 , \quad {\rm if}\ \ i = j \end{array} \right. ,
\end{align}
and the operator $\hat{t}_h$ reverses the sign of the magnetic field.
Acting $\Theta_x$ twice yields,
\vskip -0.5cm
\begin{align}
\Theta_x^2 \Psi_{n{\vec k}}({\vec r}) = \Psi_{n{\vec k}}(S_x^2{\vec
  r}) = \Psi_{n{\vec k}}(T_{100} {\vec r}) = e^{ik_x} \Psi_{n{\vec k}}({\vec r}) .
\end{align}\\

\noindent {\bf Refraction}\\
The photonic dispersion in the medium with refraction index $n_i$ is
given by $\ome=c|{\vec k}|/n_i$. {We consider a light beam
injected from a medium with a refraction index $n_i>1.65$ into the PhC
to enable frequency and wavevector matching with the Dirac
cones.} Around the DP at ${\vec K}_0=(\pi,\pi,k_0)$, the dispersion in
the medium can be expressed as $\ome = c|{\vec k}|/n_i= c|{\vec
  K}_0+{\vec q}|/n_i$ where ${\vec q}={\vec k}-{\vec K}_0$. Since the perpendicular
wavevector $k_z$ is not conserved during refraction, we can always set
\vskip -0.5cm
\begin{align}
q_x = q_\parallel\cos(\phi_i), \quad q_y = q_\parallel\sin(\phi_i) 
\end{align}
for fixed $q_\parallel$, while adjusting $k_z$ to keep a constant
frequency. The angle $\phi_i$ is varied from 0 to $2\pi$.
The refraction in the $x$-$z$ plane is determined by matching
the frequency and the parallel wavevector, yielding 
\vskip -0.5cm
\begin{align}
\ome - \ome_0 = v (\eta q_z + \tau\sqrt{q_z^2 + \gamma^2 q_x^2}),
\quad \tau = \pm 1
.
\end{align}
The perpendicular wavevector $q_z$ in the PhC is determined by
the above equation, which has two solutions for $\eta<-1$ 
\vskip -0.5cm
\begin{align}
q_{z}^{\tau} = \frac{\eta(\ome-\ome_0)+ \tau
  \sqrt{(\ome-\ome_0)^2+v^2(\eta^2-1)\gamma^2 q_x^2}}{v(\eta^2-1)} . \label{qzsol}
\end{align} 
{The refraction angle is determined through the group velocities in the
PhC as, $\theta_r \equiv -\arctan(\frac{v_x}{v_z})$. Using the 
dispersion in Eq.~(\ref{dpII}), we find that $v_z = v \left( \eta +
  \frac{\tau q_z}{\sqrt{\gamma^2 q_x^2 + q_z^2}} \right)$, 
$v_x = \frac{\tau v \gamma q_x}{\sqrt{\gamma^2 q_x^2 +
    q_z^2}}$.} Inserting Eq.~(\ref{qzsol}) into the definition of the
refraction angle, we obtain
\vskip -0.5cm
\begin{align}
\theta_{r2}=-\theta_{r1} = - \arctan\left(\frac{v \gamma
    q_x}{\sqrt{(\ome-\ome_0)^2+v^2(\eta^2-1)\gamma^2 q_x^2}}\right).
\end{align}
Refraction for generic ${\vec q}$ (i.e., away from the $x$-$z$ or
$y$-$z$ plane) is given in details in the Supplementary Materials.\\

\noindent {\bf Calculation of surface states}\\
The surface states are obtained by supercell calculations. The
supercell is periodic in the $y$-$z$ plane but finite in the $x$
direction. There are seven layers of unit cell along this direction as
sandwiched by air layers of length $3a$ on the left and right,
separately. The simple cladding medium (air) used here is
non-topological for all polarizations and useful to study topological
surface states below the light-line. The supercell structure is set to preserve the $S_y$
symmetry. Since $S_y$ transforms $(k_y,k_z)$ to
$(k_y,-k_z)$ in the surface BZ, the surface spectrum is symmetric with
respect to $k_z=0$ and $k_z=\pi$. In addition, the ${\cal T}$ symmetry
guarantees that the surface spectrum is invariant under the
transformation $(k_y,k_z)$ to $(-k_y,-k_z)$. Therefore
the surface photonic dispersion is also symmetric with respect to
$k_y=0$ and $k_y=\pi$. {As detailed in
  Ref.~\cite{arxiv}, although there are other topological degeneracies 
  in our PhC (such as nodal lines), they do not affect the surface states on
  the (100) and (010) surfaces.}

In the calculation of the reference slab-defect states, we have set
the permittivity of the slab-defect layer as $\vep=8$ (the same as
that of the dielectric slab on top of the topological PhC). The logs
of the woodpile PhCs above and below the slab-defect layer are of
width 0.25$a$, height $0.3a$ and permittivity of 12 (silicon).\\

\noindent {\bf \Large Acknowledgements}\\
We thank Sajeev John, Zhengyou Liu, Ling Lu, Chen Fang, Huanyang
Chen, Yun Lai, Chunying Qiu, and Jie Luo for many inspiring
discussions. \\

\noindent {\bf \Large Competing interests}\\
The authors declare no competing financial interests.\\

\noindent {\bf \Large Contributions}\\
J.H.J conceived the idea and wrote the manuscript. J.H.J and Z.H.H
designed the photonic architecture. H.X.W, Y.C, H.Y.K and J.H.J did the
theoretical analysis and calculations. J.H.J guided the research. \\

\noindent {\bf \Large Funding}\\
H.X.W and J.H.J acknowledge supports from the National Science
Foundation of China (Grant no. 11675116) and the Soochow university. Z.H.H
is supported by National Science Foundation of China (Grant
no. 11574226). Y.C and H.Y.K are supported by NSERC of 
Canada and Center for Quantum Materials at the University of Toronto.\\

\noindent {\bf \Large Data availability}\\
{All relevant data are available from the corresponding author J.H.J
(email: jianhuajiang@suda.edu.cn or joejhjiang@hotmail.com).}\\

\noindent {\bf \Large References}\\

{}

\vskip 5cm

\noindent {\bf \Large Figure Legends}\\

\begin{figure}
\begin{center}
\includegraphics[width=8cm]{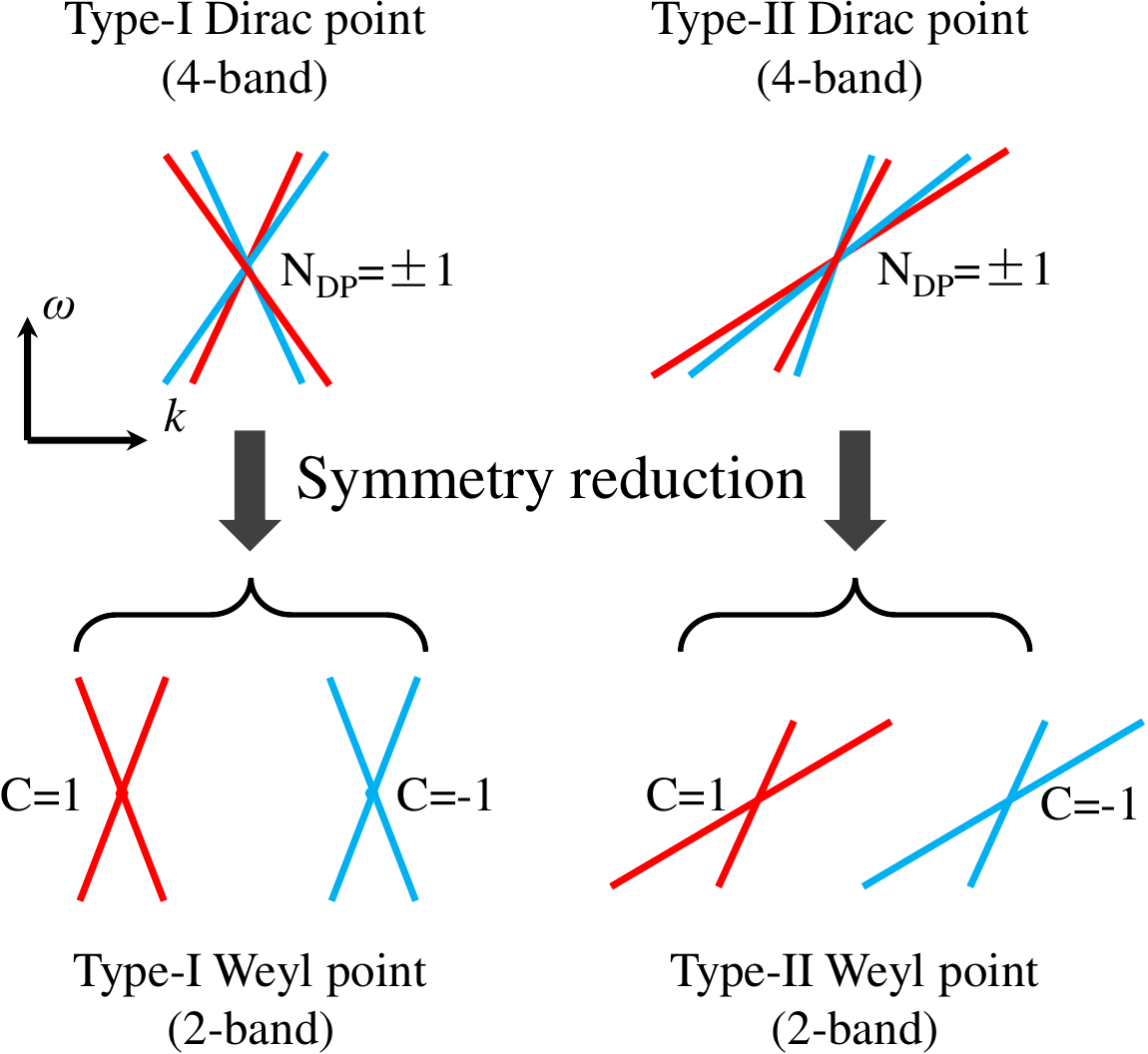}
\caption{ {\bf Type-I and type-II Dirac/Weyl points}. A type-I Dirac
  point (4-band degeneracy point), characterized by a topological
  number $N_{DP}=\pm 1$, consists of two type-I Weyl points (2-band
  degeneracy points) with opposite Chern number, $C=\pm 1$. A type-II
  Dirac point consists of two type-II Weyl points with Chern number
  $C=\pm 1$. A type-I Dirac (Weyl) point has four (two) branches,
  among which there are both positive and negative group velocities.
  In a type-II Dirac/Weyl point, instead, there are only branches of
  positive (or negative, not shown in the figure) group
  velocities. The definition of the topological charge of the Dirac
  points $N_{DP}$ are given in the main text.}
\end{center}
\end{figure}

\begin{figure}
\begin{center}
\includegraphics[width=16cm]{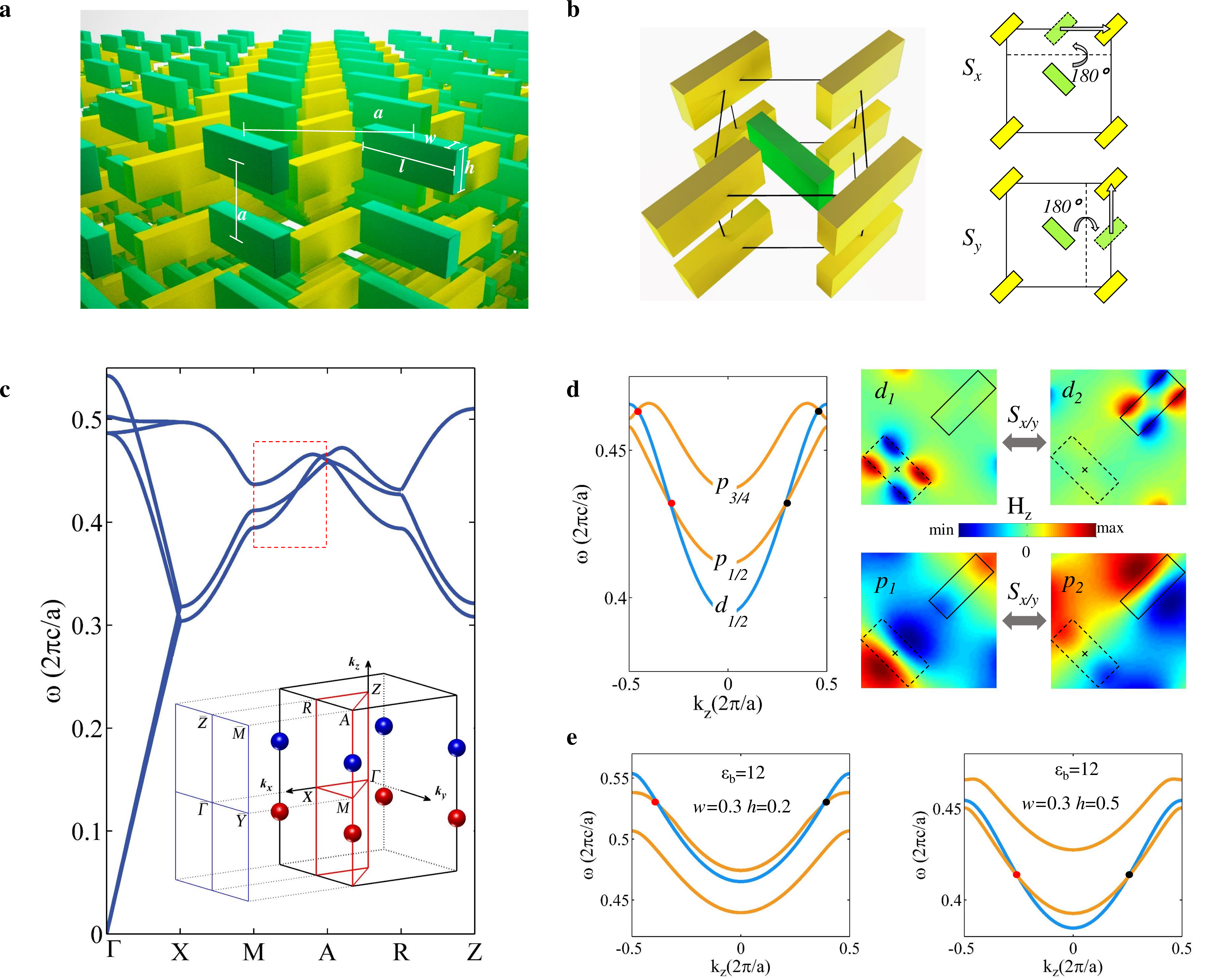}
\caption{ {\bf All-dielectric photonic-crystals for Dirac points}. {\bf
    a}, 3D view of the PhC. The yellow and green blocks are of the
  same material and shape and permittivity $\vep_b$. The background
  is polymer $\vep_m=1.9$. {\bf b}, Left: 3D structure of a unit cell
  (boundaries are indicated by black lines). The lattice constant
  along all directions is $a$. Right: Illustration of the two
  orthogonal screw symmetries $S_x$ and $S_y$ in top-down view. {\bf
    c}, Photonic band structure for $\vep_b=16$, $l=0.5a$, $w=0.2a$,
  and $h=0.5a$. Inset: bulk and surface BZs. The $k_x=\pi$ plane is
  doubly degenerate due to the screw symmetry. {\bf d}, Left: Parity
  inversion on the MA line (each curve represents a 
  doublet). Right: magnetic field profiles of the $p$- and $d$-wave
  doublets, $p_{1/2}$ and $d_{1/2}$, respectively. The doubly
  degenerate states are connected by the screw symmetries. {\bf e},
  Band inversion for other substantially different parameters. 
  DPs of topological charge $N_{DP}=+1, -1$ are
  labeled as red and blue (red and black) in {\bf c} ({\bf e}),
  separately. Frequencies are in unit of $2\pi c/a$ with $c$ being the
  speed of light in vacuum.}
\end{center}
\end{figure}

\begin{figure}
\begin{center}
\includegraphics[width=14cm]{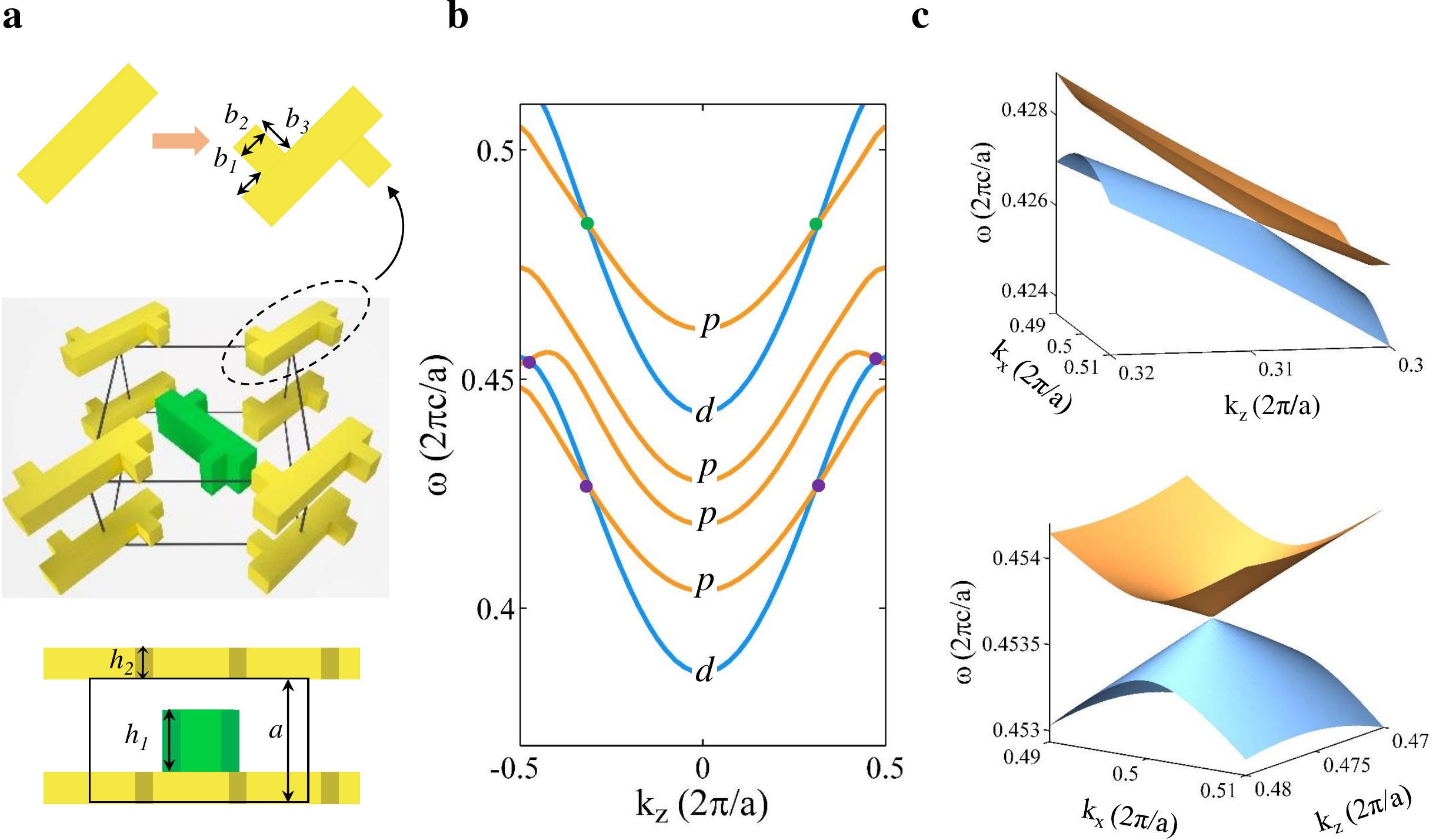}
\caption{ {\bf Weyl points derived from Dirac points.} {\bf a},
  Unit-cell structure of the symmetry-broken PhC. Upper panel: 3D
  view with zoom-in illustration of structure deformations. Lower
  panel: side view from [1$\bar{1}$0] direction. 
  The geometry parameters are $b_1=0.1$,
  $b_2=0.11$, $b_3=0.094$, $h_1=0.5$, and $h_2=0.3$. The $z$
  coordinates of the centers of the two types of dielectric blocks 
  are 0 and $0.65a$, respectively. {\bf b}, Band structure on the MA
  line for part of the first six photonic bands indicates removal of double
  degeneracy and linear-crossing between non-degenerate $p$- and
  $d$-states. These crossings are identified as type-I and type-II
  WPs. Purple (green) dots stand for WPs with Chern number -1
  (+1). {\bf c}, Dispersions of a type-II WP (upper panel) and a
  type-I WP (lower panel). The former is due to the crossing between
  band 1 and 2, while the latter originates from the crossing between
  band 1 and 3. The bands are numerated in ascending order at the M
  point.}
\end{center}
\end{figure}

\begin{figure}
\begin{center}
\includegraphics[width=11cm]{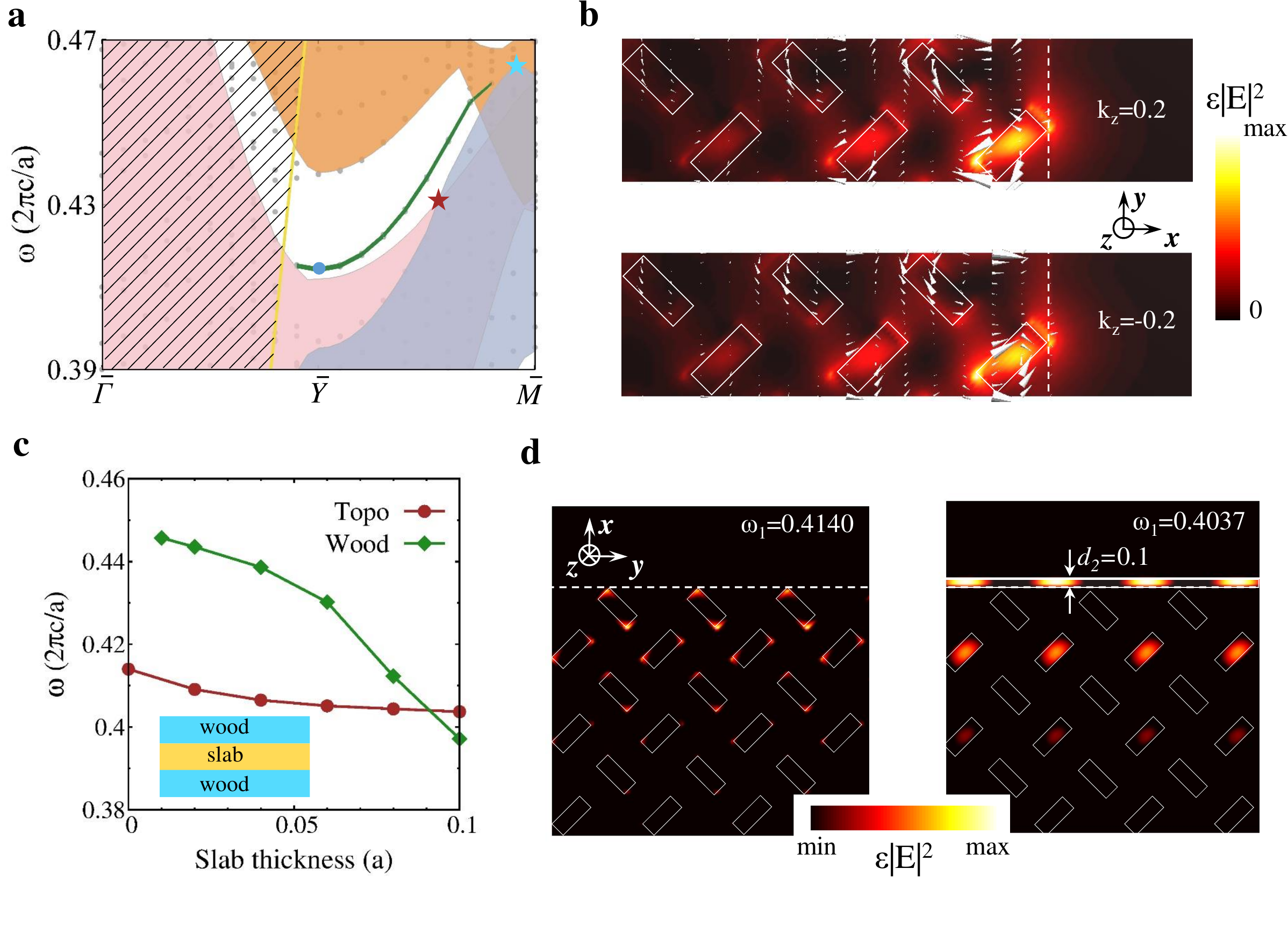}
\caption{ {\bf Dirac points and topological surface states.} {\bf a},
  Surface band and projected bulk bands on (100)
  PhC-air interface. The blue (brown) star represents a type-I
  (type-II) DP, which are the crossing point of band 1,2 
  (gray region) and band 5,6 (brown region) [band 3,4
  (pink region)]. The surface band (green curve) is below
  the light-line (the golden line). The region above the light-line is
  depicted by the shadow. Gray dots represent the spectrum from a
  finite-size supercell calculation (see Methods). {\bf b}, Energy
  density and Poynting vector profiles for the topological surface
  states at two opposite wavevectors with $k_y=\pi$.
  {\bf c}, Stability of the topological (``Topo'') surface
  states. Frequency of the topological surface state at $\bar{Y}$
  vs. thickness of a slab with permittivity $\vep=8$ on top of the PhC
  surface. The reference curve is the same dependence for the
  slab-defect state induced by a slab of the same permittivity embedded in a
  woodpile (``Wood'') PhC (structure schematically shown in the
  inset). {\bf d}, Field energy distribution of the topological surface
  states at $\bar{Y}$ point for slab
  thickness 0 (Left) and $0.1a$ (Right), respectively.}
\end{center}
\end{figure}

\begin{center}
  \begin{figure}
  \includegraphics[width=15cm]{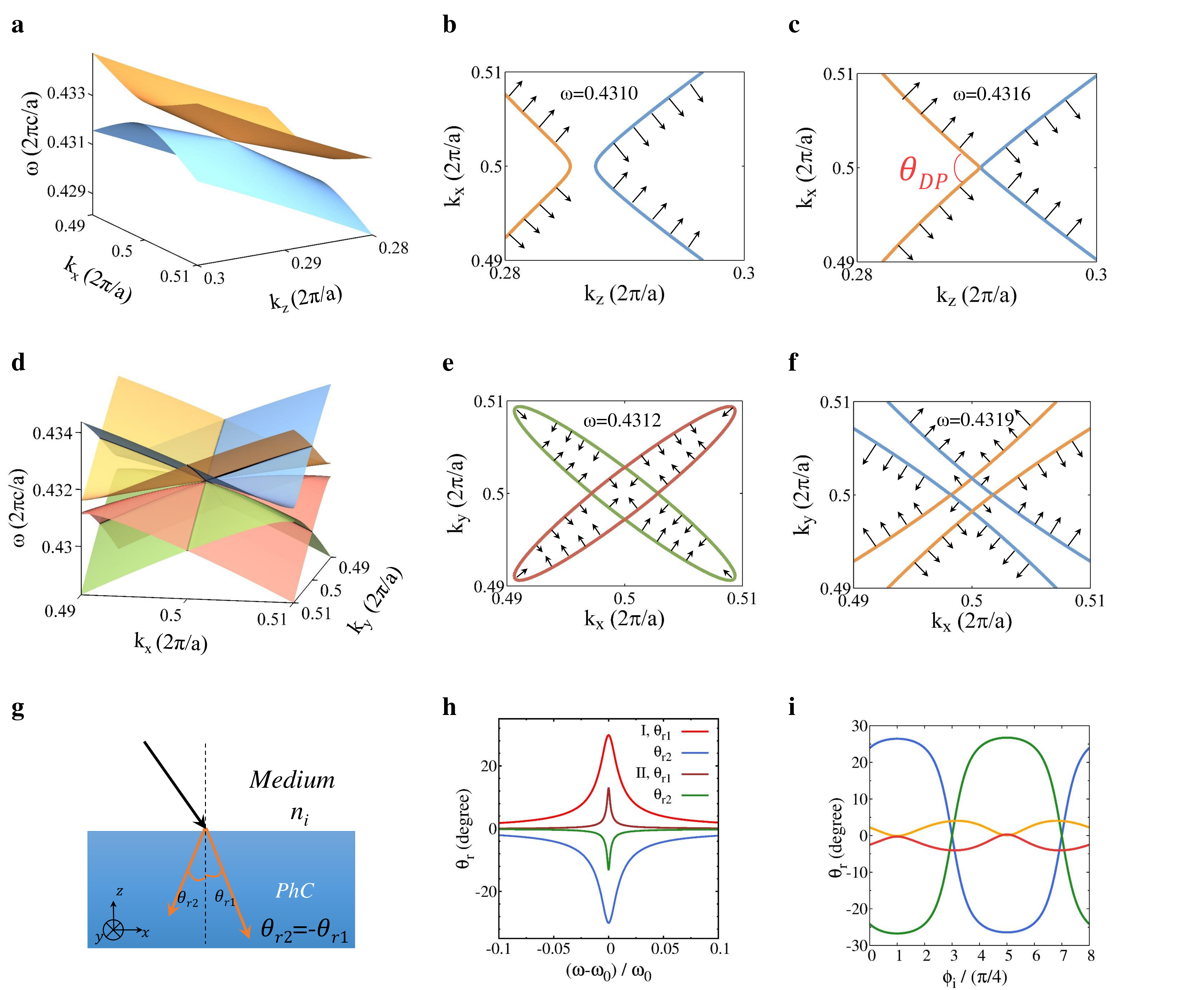}
    \caption{ {\bf Type-II Dirac cone and anomalous refraction}. {\bf
        a}, Type-II Dirac dispersion on the $k_x$-$k_z$ plane. {\bf
        b}, Isofrequency contours (orange and blue curves for the
      upper and lower branches, respectively) and group velocities
      (black arrows) on the $k_x$-$k_z$ plane for a frequency below
      the type-II DP. {\bf c}, Similar to {\bf b}, but for the
      frequency at the DP. {\bf d}, Dispersion of the Dirac cone on the
      $k_x$-$k_y$ plane. {\bf e} and {\bf f}, Isofrequency contours on
      the $k_x$-$k_y$ plane for frequencies below and above the DP,
      respectively. {\bf g}, Schematic of anomalous refraction of
      type-II DPs: concurrent positive and
      negative refraction with opposite angles. {\bf h}, Refraction
      angles vs. frequency for two cases with $\phi_i=0$. Case I:
      $\gamma=1$ and $q_\parallel=0.2\frac{\pi}{a}$. Case
      II: $\gamma=0.4$ and $q_\parallel=0.1\frac{\pi}{a}$. {\bf i}, Refraction angles
      vs. angle $\phi_i$ for $\gamma=1$ and $q_\parallel=0.1\frac{\pi}{a}$ and
      $\frac{\delta\ome}{\ome_0}=0.005$ with $\ome_0=0.4\frac{2\pi c}{a}$.}
\end{figure}
\end{center}


\begin{thebibliography}{999}

\bibitem{dirac} Dirac, P. A. M. The quantum theory
  of the electron. {\it Proc. Roy. Soc. A} (London) {\bf 117}, 610-624. (1928). 

\bibitem{ti1} Hasan, M. Z. \& Kane, C. L. Topological insulators.
  {\it Rev. Mod. Phys.} {\bf 82}, 3045-3067 (2010).

\bibitem{ti2} Qi, X.-L. \& Zhang, S.-C. Topological insulators and
  superconductors. {\it Rev. Mod. Phys.} {\bf 83}, 1057-1110 (2011).

\bibitem{tsm1} Vafek, O. \& Vishwanath, A. Dirac fermions in
    solids: from high-$T_c$ cuprates and graphene to topological
    insulators and Weyl semimetals.
  {\it Ann. Rev. Cond. Matt. Phys.} {\bf 5}, 83-112 (2014).


\bibitem{haldane} Haldane, F. D. M. \& Raghu, S. Possible
  realization of directional optical waveguides in photonic crystals
  with broken time-reversal symmetry. 
  {\it Phys. Rev. Lett.} {\bf 100}, 013904 (2008).


\bibitem{rev1} Lu, L., Joannopoulos, J. D. \& Solja\v{c}i\'{c}, M.
  Topological photonics.
  {\it Nat. Photon.} {\bf 8}, 821-829 (2014).


\bibitem{bzhang} Yang, Z., Gao, F., Shi, X., Lin, X., Gao, Z., Chong, Y. \&
  Zhang, B. Topological acoustics. {\it Phys. Rev. Lett.} {\bf 114}, 114301 (2015).

\bibitem{acoustic} Xiao, M., Chen, W.-J., He, W.-Y. \&
  Chan, C. T. Synthetic gauge flux and Weyl points in acoustic
    systems. {\it Nat. Phys.} {\bf 11}, 920-924 (2015). 

 \bibitem{nju1} He, C., Ni, X., Ge, H., Sun, X.-C., Chen, Y.-B., Lu,
   M.-H., Liu, X.-P. \& Chen, Y.-F. Acoustic topological insulator and
     robust one-way sound transport. {\it Nat. Phys.} {\bf 12}, 1124-1129
   (2016).

\bibitem{huber} S\"usstrunk, R. \& Huber, S. D. Observation of phononic helical edge states in a
  mechanical topological insulator. {\it Science} {\bf 349}, 47-50 (2015).

\bibitem{vinzo} Rocklin, D. Z., Chen, B. G.-g., Falk, M., Vitelli,
  V. \& Lubensky, T. €‰C. 
  Mechanical Weyl modes in topological Maxwell lattices.
  {\it Phys. Rev. Lett.} {\bf 116}, 135503 (2016).

\bibitem{ZB} Zhang, X. Observing Zitterbewegung for photons near
    the Dirac point of a two-dimensional photonic crystal. {\it Phys. Rev. Lett.} {\bf 100}, 113903 (2008).

\bibitem{zim} Huang, X. Q., Lai, Y., Hang, Z. H., Zheng, H. H. \&
  Chan, C. T. Dirac cones induced by accidental degeneracy in photonic
  crystals and zero-refractive-index materials. {\it Nat. Mater.} {\bf 10}, 582-586 (2011). 

\bibitem{sMag} Rechtsman, M. C., Zeuner, J. M., T\"unnermann, A., 
  Nolte, S., Segev, M. \& Szameit, A. Strain-induced
    pseudomagnetic field and photonic Landau levels in dielectric
    structures. {\it Nat. Photon.} {\bf 7}, 153-158
  (2013). 


\bibitem{z2meta} Khanikaev, A. B., Mousavi, S. H., Tse, W.-K., 
  Kargarian, M., MacDonald, A. H. \& Shvets, G. Photonic
    topological insulators. {\it Nat. Mater.} {\bf
    12}, 233-239 (2013).


\bibitem{ctti} Chen, W.-J., Jiang, S.-J., Chen, X.-D., 
  Dong, J.-W., \& Chan, C. T. Experimental realization of photonic
    topological insulator in a uniaxial metacrystal waveguide.
  {\it Nat. Commun.} {\bf 5}, 5782 (2014).

\bibitem{shvets} Ma, T., Khanikaev, A. B., Mousavi, S. H. \& Shvets,
  G. Guiding electromagnetic waves around sharp corners:
    topologically protected photonic transport in metawaveguides. 
  {\it Phys. Rev. Lett.} {\bf 114}, 127401 (2015).

\bibitem{huxiao} Wu, L.-H. \& Hu, X. Scheme for achieving a
    topological photonic crystal by using dielectric material.
  {\it Phys. Rev. Lett.} {\bf 114}, 223901 (2015). 

\bibitem{oe1} Xu, L., Wang, H.-X., Xu, Y.D., Chen, H.Y. \& Jiang, J.-H.,
    Accidental degeneracy in photonic bands and topological phase
       transitions in two-dimensional core-shell dielectric photonic
       crystals. 
     {\it Opt. Express} {\bf 24}, 18059-18071 (2016).


\bibitem{mit} Wang, Z., Chong, Y., Joannopoulos, J. D. \&
  Solja\v{c}i\'{c}, M. Observation of unidirectional
    backscattering-immune topological electromagnetic states.
  {\it Nature} (London) {\bf 461}, 772-775 (2009).

 \bibitem{wu} Poo, Y., Wu, R.-X., Lin, Z., Yang, Y. \& Chan, C. T. 
     Experimental Realization of Self-Guiding Unidirectional
     Electromagnetic Edge States. 
   {\it Phys. Rev. Lett.} {\bf 106}, 093903 (2011).


\bibitem{hafezi2} Hafezi, M., Mittal, S., Fan, J., Migdall, A. \&
  Taylor, J. Imaging topological edge states in silicon photonics.
  {\it Nat. Photon.} {\bf 7}, 1001-1005 (2013).

\bibitem{floquet} Rechtsman, M. C., Zeuner, J. M., Plotnik, Y., Lumer, Y.,
  Podolsky, D., Dreisow, F., Nolte, S., Segev, M. \& Szameit, A. 
    Photonic Floquet topological insulators. 
  {\it Nature} (London) {\bf 496}, 196-200 (2013). 


\bibitem{hafezi3} Mittal, S., Ganeshan, S., Fan, J.,  Vaezi, A. \&
  Hafezi, M. Measurement of topological invariants in a 2D photonic
  system. {\it Nat. Photon.} {\bf 10}, 180-183 (2016).


\bibitem{ling1} Lu, L., Fu, L., Joannopoulos, J. D.
  \& Solja\v{c}i\'{c}, M. Weyl points and line nodes in gyroid
    photonic crystals. {\it Nat. Photon.} {\bf 7}, 294-299 (2013).

\bibitem{ling-exp} Lu, L., Wang, Z., Ye, D., Ran, L., Fu, L.,
  Joannopoulos, J. D. \& Solja\v{c}i\'{c}, M. Experimental
    observation of Weyl points.
  {\it Science} {\bf 349}, 622-624 (2015).

\bibitem{szhang} Gao, W., Yang, B., Lawrence, M., Fang, F., B\'eri, B. \&
   Zhang, S. Plasmon Weyl degeneracies in magnetized plasma.
  {\it Nat. Comm.} {\bf 7}, 12435 (2016).

\bibitem{ct-exp} Chen, W.-J., Xiao, M. \& Chan, C. T. Experimental
  observation of robust surface states on photonic crystals possessing
  single and double Weyl points. {\it Nat. Commun.} {\bf 7}, 13038 (2016).

\bibitem{3ddp} Wang, H.-X., Xu, L., Chen, H.Y. \& Jiang, J.-H. 
    Three-dimensional photonic Dirac points stabilized by point group
    symmetry. {\it Phys. Rev. B} {\bf 93}, 235155 (2016). 

\bibitem{xiao} Xiao, M., Lin, Q. \& Fan, S. Hyperbolic Weyl point
    in reciprocal chiral metamaterials. {\it Phys. Rev. Lett.} {\bf 117}, 057401 (2016).

\bibitem{3dti} Lu, L., Fang, C., Fu, L., Johnson, S. G., 
  Joannopoulos, J. D. \& Solja\v{c}i\'{c}, M. Symmetry-protected
    topological photonic crystal in three dimensions. 
  {\it Nat. Phys.} {\bf 12}, 337-340 (2016).


\bibitem{3dwti} Slobozhanyuk, A., Mousavi, S. H., Ni, X.,
  Smirnova, D., Kivshar, Y. S. \& Khanikaev, A. B. 
    Three-dimensional all-dielectric photonic topological insulator.
  {\it Nat. Photon.} {\bf 11}, 130-136 (2017).

\bibitem{fukane} Fu, L. \& Kane, C. L. Topological insulators with inversion symmetry. {\it Phys. Rev. B} {\bf 76}, 045302 (2007).


\bibitem{WPII} Soluyanov, A. A., Gresch, D., Wang, Z., Wu, Q., Troyer, M., 
  Dai, X. \& Bernevig, B. A. Type-II Weyl semimetals. {\it Nature}
  (London) {\bf 527}, 495-498 (2015).

\bibitem{WP2} Xu, Y., Zhang, F. \& Zhang, C. Structured Weyl points in spin-orbit coupled fermionic superfluids.
  {\it Phys. Rev. Lett.} {\bf 115}, 265304 (2015).

 \bibitem{mermin} K\"onig, A. \& Mermin, N. D. Electronic level
     degeneracy in nonsymmorphic periodic or aperiodic crystals. 
   {\it Phys. Rev. B} {\bf 56}, 13607-13610 (1997).

 \bibitem{ashvin} Parameswaran, S. A., Turner, A. M., Arovas,
   D. P. \& Vishwanath, A. Topological order and absence of band
   insulators at integer filling in non-symmorphic crystals.
 {\it Nat. Phys.} {\bf 9}, 299-"303 (2013).


\bibitem{ald} Lee, J.-H., Leung, W., Ahn, J., Lee, T., 
  Park, I.-S., Constant, K. \& Ho, K.-M. {\it Layer-by-layer
    photonic crystal fabricated by low-temperature atomic layer
    deposition}. Appl. Phys. Lett. {\bf 90}, 151101 (2007).


\bibitem{dlw} Deubel, M., von Freymann, G., Wegener, M., Pereira, S.,
  Busch, K. \& Soukoulis, C. M. Direct laser writing of three-dimensional
    photonic-crystal templates for telecommunications. {\it Nat.
  Mater.} {\bf 3}, 444-447 (2004). 


\bibitem{mpb} http://ab-initio.mit.edu/wiki/index.php/MIT$_{}$Photonic$_{}$Bands.

\bibitem{yang} Yang, B.-J., Morimoto, T. \& Furusaki, A.
  Topological charges of three-dimensional Dirac semimetals with
  rotation symmetry. {\it Phys. Rev. B} {\bf 92}, 165120 (2015).

\bibitem{furusaki} Morimoto, T. \& Furusaki, A. Weyl and Dirac
    semimetals with $Z_2$ topological charge. {\it Phys. Rev. B} {\bf 89}, 235127 (2014).



\bibitem{sjyang} John, S. \& Yang, S. Electromagnetically
    Induced Exciton Mobility in a Photonic Band Gap.
  {\it Phys. Rev. Lett.} {\bf 99}, 046801 (2007).

\bibitem{noda2} Ishizaki, K. \& Noda, S. Manipulation of photons
     at the surface of three-dimensional photonic crystals. {\it Nature} (London)
   {\bf 460}, 367-370 (2009).


\bibitem{book} Joannopoulos, J. D., Johnson, S. G., Winn, J. N. \&
  Meade, R. D. {\it Photonic Crystals: Molding the Flow of Light}.
  (Princeton University Press, 2008).


\bibitem{luo} Luo, J., Xu, P., Sun, T. \& Gao, L. Tunable beam splitting and
  negative refraction in heterostructure with metamaterial.
{\it Appl. Phys. A} {\bf 104}, 1137-1142 (2011).

\bibitem{tci} Fu, L. Topological crystalline insulators.
  {\it Phys. Rev. Lett.} {\bf 106}, 106802 (2011).


\bibitem{ave} Fu, L. \& Kane, C. L. Topology, delocalization via
    average symmetry and the symplectic Anderson transition.
  {\it Phys. Rev. Lett.} {\bf 109}, 246605 (2012).


\bibitem{hasan} Chang, T.-R. {\sl et al.} Type-II symmetry-protected topological Dirac semimetals.
  {\it Phys. Rev. Lett.} {\bf 119}, 026404 (2017).


 \bibitem{arxiv} Wang, H.-X., Chen, Y., Hang, Z. H., Kee, H.-Y. \&
   Jiang, J.-H. 3D $Z_2$ Topological Nodes in Nonsymmorphic
     Photonic Crystals: Ultrastrong Coupling and Anomalous Refraction.
   {Preprint at https://arxiv.org/abs/1608.02437}.


\end{thebibliography}
\end{document}